\documentclass[a4paper,twocolumn,11pt]{quantumarticle}
\pdfoutput=1
\usepackage[utf8]{inputenc}
\usepackage[english]{babel}
\usepackage[T1]{fontenc}
\usepackage{amsmath}
\usepackage{hyperref}

\usepackage{multirow}
\usepackage{longtable}

\usepackage[numbers,sort&compress]{natbib}

\usepackage{tikz}
\usetikzlibrary{arrows.meta}
\usepackage{hyperref}
\usepackage{array}

\newcolumntype{P}[1]{>{\centering\arraybackslash}p{#1}}

\begin{document}
\def \neighl #1#2{
        \node at (#1, #2-1)[circle,fill,inner sep=2pt]{};
        \draw (#1, #2-1) -- (#1, #2+0.3);
        \draw (#1, #2) circle (0.3);
        \node[draw, fill=white] at (#1+1.5, #2) {R$_Z$($\gamma_{i,i-1}$)};
        \node at (#1+3, #2-1)[circle,fill,inner sep=2pt]{};
        \draw (#1+3, #2-1) -- (#1+3, #2+0.3);
        \draw (#1+3, #2) circle (0.3);

}
\def \neighr #1#2{
        \node at (#1, #2+1)[circle,fill,inner sep=2pt]{};
        \draw (#1, #2+1) -- (#1, #2-0.3);
        \draw (#1, #2) circle (0.3);
        \node[draw, fill=white] at (#1+1.5, #2) {R$_Z$($\gamma_{i,i+1}$)};
        \node at (#1+3, #2+1)[circle,fill,inner sep=2pt]{};
        \draw (#1+3, #2+1) -- (#1+3, #2-0.3);
        \draw (#1+3, #2) circle (0.3);
}
\def \figref #1{Fig. (\ref{#1})}

\title{Searching for Possible Spin Configurations of Ferrum Chain via Quantum Approximate Optimization Algorithm}

\author{Saba Arife Bozpolat}
\affiliation{Physics Department, Marmara University, 34722 Ziverbey, Istanbul, Turkey}
\orcid{0000-0002-2166-6464}
\homepage{saba.bozpolat@gmail.com,\\OrcID: 0000-0002-2166-6464}

\maketitle

\begin{abstract}
Calculating the expected spin configuration of the chain consisting of Ferrum atoms interacting with each other through exchange interaction is fundamentally a configuration optimization problem.
Quantum Approximate Optimization Algorithm is a suitable candidate to configure such systems on a quantum device. In this work we have considered Ferrum chains of three different lengths and calculated their most-probable spin configurations using Quantum Approximate Optimization Algorithm. We employed a Quantum Feed Forward Neural Network as the optimizer of Quantum Approximate Optimization Algorithm. We have successfully obtained the expected spin configuration for the longest Ferrum Chain.

\end{abstract}

\section{Introduction\label{sec:intro}}

In Physics, the Ising model is a well established theoretical framework to describe spin orientations of magnetic members of a magnetic material, among some other things, considering their interactions with one another and some other internal or external physical entities \citep{ising1925}.
This model is also used interdisciplinarily to build Quadratic Unconstrained Binary Optimization (QUBO) models \citep{kochenberger2014, kochenberger2018, andrew2014} to solve several Combinatorial Optimization Problems (COP) \citep{korte2006}, such as traveling salesperson problem (TSP) \citep{applegate2006}, workflow scheduling (WS) \citep{kousalya2017}, the intersection traffic signal control problem (ITSCP) \citep{eom2020}, etc. 
QUBO helps with the construction of an enunciative square matrix for considered problem. 
And this matrix is then subjected to an optimization model to fine-tune the problem-specific parameters. 
Quantum Approximate Optimization Algorithm is a quantum algorithm to do such calibrations in polynomial time whereas its classical counterparts cannot provide any guarantee on the time scale \citep{farhi2014a, farhi2014b, barak2015, ibm.qaoa}. 

Much as coupling QAOA with classical optimizers to apply this method on some COPs is a relatively novel and beneficial idea, it is possible to take one step further and integrate it with Deep Neural Networks (DNN) to capitalize swiftness of pioneering Machine Learning (ML) algorithms.

In this work, we obtained the optimized configuration of a ferromagnetic (FM) material structured in 1 dimension, $i.e.$ a Ferrum chain, without an external magnetic field, using QAOA with Quantum Feed Forward Neural Network (QFFNN). We have considered a Ferrum chain of three different lengths and expressed this sistem using Heisenberg Hamiltonian with nearest-neighbor Ising Model. We have converted this Hamiltonian to a suitable unitary gate representation to build a suitable quantum circuit. We have implemented this circuit on a quantum computer simulator using Cirq \citep{cirq}, TensorFlow \citep{tf} and TensorFlow Quantum \citep{tfq}. By using QAOA with QFFNN as the optimizer to configure a Ferrum chain under no influence of external field of any kind, We have observed that the most likely configurations are the most organized ones consistent with energy minimization.

\section{Material and Method}
\label{sec:ferrum}

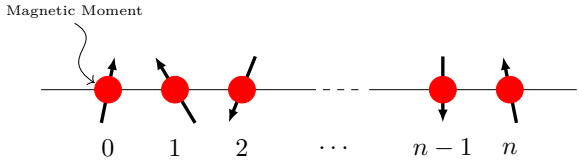
\begin{figure}[tb]
    \centering
    \resizebox{0.95\columnwidth}{!}{%
        \begin{tikzpicture}
            \draw (0.5, 1) node[yshift=5] {\tiny Magnetic Moment};
            \draw[->] (0.5,1) .. controls (1,0.7) and (0.2,0.5) .. (0.8,0.1);
            \draw (0, 0) -- (4, 0);
            \draw[-{Latex[length=2mm]}, line width=0.5mm] (0.9,-0.5) -- (1.1,0.5);
            \draw[-{Latex[length=2mm]}, line width=0.5mm] (2.3,-0.5) -- (1.7,0.5);
            \draw[-{Latex[length=2mm]}, line width=0.5mm] (3.2,0.5) -- (2.8,-0.5);
            \foreach \i in {0,...,2}
            {
                \draw[red,fill=red] (\i+1, 0) node[black, yshift=-25] {$\i$} circle (0.2);
            }
            \draw[dashed] (4, 0) -- (4.8, 0);
            \draw (4.4, 0) node[yshift=-25] {$\dots$};
            \draw (4.9, 0) -- (8, 0);
            \draw[-{Latex[length=2mm]}, line width=0.5mm] (6,0.5) -- (6,-0.5);
            \draw[-{Latex[length=2mm]}, line width=0.5mm] (7.1,-0.5) -- (6.9,0.5);
            \foreach \i in {6,...,7}
            {
                \draw[red,fill=red] (\i, 0) circle (0.2);
            }
            \draw (6, 0)  node[black, yshift=-25] {$n-1$};
            \draw (7, 0)  node[black, yshift=-25] {$n$};
        \end{tikzpicture}
    }
    \caption{[colored online] Illustration of a chain of $n$ magnetic moments. In this work we consider three different Ferrum chains with a periodic boundary condition. Initially, Ferrum atoms have random spin orientations denoted by thick-black arrows here. Length of these arrows has no physical meaning. See the text for details.}
    \label{fig:chain}
\end{figure}

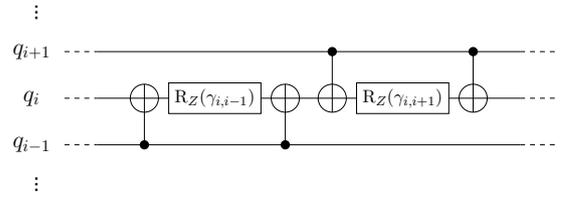
\begin{figure}[tb]
    \centering
    \resizebox{0.9\columnwidth}{!}{%
        \begin{tikzpicture}
            \draw[dashed] (1.3, 0) node[xshift=-0.7cm] {\Large $q_{i-1}$} -- (2, 0);
            \draw[dashed] (1.3, 1) node[xshift=-0.7cm] {\Large $q_i$} -- (2, 1);
            \draw[dashed] (1.3, 2) node[xshift=-0.7cm] {\Large $q_{i+1}$} -- (2, 2);
            \draw (2, 2) -- (11, 2);
            \draw (2, 0) -- (11, 0);
            \draw (2, 1) -- (11, 1);
            \draw[dashed] (11, 2) -- (11.8, 2);
            \draw[dashed] (11, 0) -- (11.8, 0);
            \draw[dashed] (11, 1) -- (11.8, 1);
            \neighl{3}{1}
            \neighr{7}{1}
            \draw[dotted, line width=0.5mm] (0.7, -0.7) -- (0.7, -1.0);
            \draw[dotted, line width=0.5mm] (0.7, 2.7) -- (0.7, 3.0);
        \end{tikzpicture}
    }
    \caption{Quantum circuit part of exchange interaction for atom on $i$-th index on the lattice. Only nearest neighbors are considered.}
    \label{fig:circuit_part_for_ith_atom}
\end{figure}

The simplest way to organize some number of magnetic moments is to place them as so that they shape a line with equidistant between each successive magnetic moments. 
This geometric arrangement is called as a chain of magnetic moments and is depicted in \figref{fig:chain}.
The Hamiltonian of a magnetic moment chain can be described by Ising Model with a Hamiltonian such as;
\begin{equation}
    H=-\sum_{i=0}^n\sum_{j=i\pm 1}J_{ij} S_i S_j,
    \label{eqn:ising_hamiltonian}
\end{equation}
where $n+1$ is total number of magnetic moments in the chain, 
\(S_i\in\{-1, 1\}\) is the spin momentum quantum number of the magnetic moment placed on the vertex \(i\),
and \(j=i\pm 1\) is the index to denote nearest neighbors of \(i\).
Here \(J_{ij}\) is the exchange coupling constant between the magnetic moments on vertices $i$ and $j$. 
Since the spin momentum quantum numbers fit the eigenvalues of Pauli-$Z$ matrix, we may represent spin-up and spin-down states of the magnetic moments with statevectors $|0\rangle$ and $|1\rangle$, respectively. 
Then, we may represent the chain's Hamiltonian using Pauli-$Z$ matrix such as,
\begin{equation}
    \hat{H}=-\sum_{i=0}^n\sum_{j=i\pm 1}J_{ij} Z_i\otimes Z_j.
    \label{eqn:ising_hamiltonian_for_ferrum}
\end{equation}
In order to write the unitary operator corresponding to this Hamiltonian, it is necessary to raise it to exponential. Therefore we get the unitary operator as,

\begin{align}
    & U(\gamma) = \prod_{i=0}^n\prod_{j=i\pm 1} U_{ij}(\gamma),
    \label{eqn:unitary_operator_for_system} 
\end{align}
where $ U_{ij}(\gamma_{ij})\equiv\exp(-i\gamma_{ij} J_{ij} Z_i \otimes Z_j)$ is the representative part of the unitary operator for the magnetic moment on the vertex-$i$.
This unitary operator is a parametrized one and it needs to be optimized to yield the most probable combinations of considered system. 
Innately, QAOA can be used to optimize these types of operators, and to do so a suitable quantum circuit needs to be configured. 
Note that this operator contains only second order $Z$ matrices and these matrices can be realized in a quantum circuit by using an $R_z(\gamma)$ gate sandwiched by two Controlled-X gates \citep{zawalska2020}. 
This way, the circuit representation for the $i$-th magnetic moment in the chain interacting with only its nearest neighbors is pictured in \figref{fig:circuit_part_for_ith_atom}. 

\begin{figure*}[t]
    \centering
    \includegraphics[width=\textwidth]{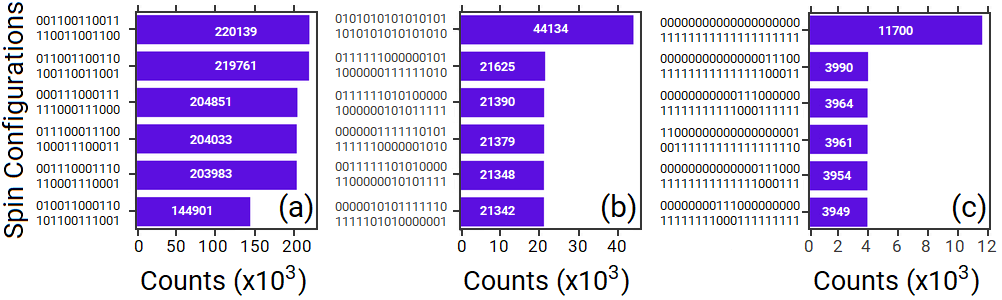}
    \caption{Counts of most-encountered spin configurations for ferrum chain of length (a) 12, (b) 16 and (c) 20 atoms.}
    \label{fig:all_lengths_tiny}
\end{figure*}

We examined three different Ferrum chains of different lengths in our simulations.
Since we map the qubit states $|0\rangle$ and $|1\rangle$ to up and down spin orientations of Ferrum atoms, respectively, n+1 qubits should be employed in the quantum circuit in order to represent the entire spin configuration space of the Ferrum chain with n+1 Ferrum atoms.  
The qubits were initially put into an equi-superposed state.
Then, we applied QAOA to this system by applying a phase ten times in which the two operators are applied alternately; first one is the problem unitary operator given in Eqn. \ref{eqn:unitary_operator_for_system} and the second one is a mixing unitary operator selected as $X$ gate \citep{hadfield2019} applied on each qubit on the quantum circuit.
Therefore we introduced 20 input parameters into the circuit and initial values of these parameters are randomly assigned between 0 and $2\pi$ radians.
All of the qubits in the circuit was measured at the end of the circuit execution and this measurement yielded an (n+1)-bit string which represents the spin configuration of the Ferrum chain.
The energy eigenvalue corresponding to the obtained spin configuration was used to optimize the circuit parameters via QFFNN. Adam optimizer \citep{kingma2014} was used for optimization with 125 epoches.
After training of QFFNN, we sampled 50 million spin configurations and counted the number of different spin configurations occurred during the sampling.
To imitate a physically infinite chain in the calculations, we have applied the periodic boundary conditions \citep{rapaport_2004}. 
In literature, \(J_{ij}\) is taken as \(J_{soft}\), or \(J_s\) for short, for Ferrum atoms and its value is 1.5 peV/m for a Ferrum chain \citep{deuger2016}.

The optimized spin configuration of such systems is either all-spin-up (``$111\dots11$'') or all-spin-down (``$000\dots00$'') both of which has the same energy value.
This orientation situation is independent of the chain length and only due to the nature of ferromagnetic materials.
On the other hand, dimension of the spin configuration space increases exponentially with the number of atoms in the chain. 
This actuality yields a rapid decrease on the probability of getting one of the optimized spin configurations in a search by brute force. 
However, our solution finds the possible configuration with increasing reliability for increasing chain length.

\section{Results}
\label{sec:results}

After analysing the data, we have counted the most-encountered spin configurations obtained from sampling via optimized parameters. 
Since there is no external magnetic field is introduced to the system, all-spin-up and all-spin-down configurations are essentially the same. Therefore we have counted a spin configuration along with its upside-down version.
First a few most captured spin configurations for Ferrum chains of length 12, 16 and 20 atoms are showed in \figref{fig:all_lengths_tiny}. 

Dimension of the configuration space increases exponentially with the number of atoms in the chain. 
Notwithstanding, there are only two expected configurations. 
This situation is independent of the chain length and only due to the nature of ferromagnetic materials. 
This characteristic yields a rapid decrease on the probability of getting the right solution, if the ordered seek algorithms would be employed. 
The probability of getting one of two all-aligned spin configurations is 1 in 2048 for the chain of 12 atoms, 1 in 32768 for the chain of 16 atoms and 1 in 524288 for for the chain of 20 atoms. 
However, our solution finds the theoretically expected spin configuration with increasing reliability for increasing chain length.

\section{Conclusion}
\label{sec:conclusion}
In conclusion, we have investigated the spin configuration of a ferromagnetic chain using cutting-edge quantum computation and quantum machine learning techniques. 
In our calculations we have applied QAOA to the Ferrum chain of 3 different lengths with a quantum feed-forward neural-network as optimizer. 
We have expressed this system via Ising Hamiltonian consists of only exchange term. 
We haven't considered an external magnetic field or any other interactions between the magnetic moments of any sort.
In case of the Ferrum chain with 12 and 16 atoms, our algorithm does not yield the anticipated spin configuration. 
But when the Ferrum chain was made long enough, our algorithm was able to calculate the expected spin configuration.
This can be explained by the size effect.

\section{Acknowledgement}
\label{sec:ack}
This study was carried out within the scope of the \href{https://qworld.net/qcourse570-1/}{QC570-1} course given in partnership of \href{https://www.qworld.net}{QWorld} and \href{https://www.df.lu.lv/en/}{DF@LU}. Author sincerely thanks Abuzer Yakaryilmaz and course attendants for their valuable input to this work about quantum computation. Author also appreciates fruitful discussions with her colleague, Caner Deger about magnetism and magnetic materials.

\section{Code and Data Availability}
\label{sec:code}
Data, and the code from which is used to create this data, will be provided by the author upon reasonable request.

\bibliographystyle{unsrt}
\bibliography{main}

\end{document}